\documentclass[12pt,preprint]{aastex}

\newcommand{\vC}{$\check{\rm C}$}

\newcommand{\Les}{L_{\varepsilon,s}}
\newcommand{\Lec}{L_{\varepsilon,IC}}
\newcommand\Gf{\Gamma_{\rm f}}
\newcommand\Gs{\Gamma_{\rm s}}
\newcommand\Geq{\Gamma_{\rm beq}}
\newcommand\Ginf{\Gamma_{\rm b\infty}}
\newcommand{\D}{\mathrm{d}}
\slugcomment{Accepted for publication in the Astrophysical Journal}
\shorttitle{\uppercase{The $\delta$ crisis of T\lowercase{e}V blazars :
a clue for pileup EDF ?}}
\shortauthors{\uppercase{Henri\ \&\ Saug\'e}}
\begin{document}
%%
%% Title
%%
\title{%% 
 The bulk Lorentz factor crisis of T\lowercase{e}V blazars :\\
 evidence for an inhomogeneous pileup energy distribution ?  }
%%
%% Author and affiliation information.
%%
\author{Gilles Henri}
    \affil{
      Laboratoire d'Astrophysique de Grenoble, 
      Universit\'e Joseph-Fourier, BP~53, F-38041 Grenoble, France 
    }
\and
\author{Ludovic Saug\'e}
    \affil{
    Institut de Physique Nucl\'eaire de Lyon, UCBL/IN2P3-CNRS , 
    4, rue Enrico Fermi, F-69622, Villeurbanne cedex, France
    }
\email{Gilles.Henri@obs.ujf-grenoble.fr,l.sauge@ipnl.in2p3.fr}
\received{13 July 2005}\revised{14 November 2005}\accepted{18 November 2005}%%

\begin{abstract} 
    There is growing evidence that the estimations of the beaming Doppler
    factor in TeV BL Lac object based on the Self Synchrotron Compton (SSC)
    models are in strong disagreement with those deduced from the
    unification models between blazars and radio galaxies. When corrected
    from extragalactic absorption by the diffuse infrared background
    (DIrB), the SSC one-zone models require very high Lorentz factor
    (around 50) to avoid strong $\gamma-\gamma$ absorption.  However, the
    statistics on beamed \textit{vs.} unbeamed objects, as well as the
    luminosity contrast, favors much lower Lorentz factor of the order of
    3. In this paper, we show that for the special case of Markarian 501,
    the need for very high Lorentz factor is unavoidable for all one-zone
    models where all photons are assumed to be produced at  the same
    location at the same time.  Models assuming a double structure with two
    different beaming patterns can partially solve the problem of
    luminosity contrast, but we point out that they are inconsistent with
    the statistics on the number of detected TeV sources. The only way to
    solve the issue is to consider inhomogeneous models, where low energy
    and high energy photons are not produced at the same place, allowing
    for much smaller Lorentz factors. It implies that the jet is
    stratified, but also that the particle energy distribution is close to
    a monoenergetic one, and that pair production is likely to be
    significant. The implications on relativistic jet physics and particle
    acceleration mechanism are discussed. 
\end{abstract}


%% --------------------------------------------------------------------------------
   \section{Introduction}
%% --------------------------------------------------------------------------------

It is now admitted that the blazar phenomenon is due to relativistic
Doppler beaming of the non-thermal jet emission taking place in radio-loud
Active Galactic Nuclei (AGN) where the jet axis is closely aligned with the
observer's line of sight. They exhibit an important level of optical
polarization, a flat radio spectrum, a strong variability in all frequency
bands and a very broad spectral energy distribution (SED) ranging from the
radio to the extreme gamma ray band.  The SED consists typically in two
broad components. In the so-called Synchrotron-Self-Compton process (SSC)
model, the lowest energy hump is attributed to the synchrotron emission
from relativistic electrons and/or positrons, and the highest one is
attributed to the Inverse Compton (IC) process of the same charged particles
onto the synchrotron photons and/or external photons.  
%%%
The blazar class of objects includes both the \emph{Flat Spectrum Radio
Quasars} (FSRQ) and the \emph{BL Lac} sources (or Lacertids), depending
respectively on the existence or the lack of detectable emission lines in
their spectra. 
%%%
Following \citet{chiaberge00}, one can define two classes of BL Lac objects
(which are most probably two extreme cases in a continuous distribution) : the
LBL or "red" BL Lac, for which the synchrotron component peaks in far IR to
optical, and the IC component peaks in the MeV-GeV range, and the HBL or
"blue" BL Lac, for which the synchrotron component peaks in the UV-X range,
and the IC component peaks above 10 GeV.  The most extreme objects up to now
are those whose non thermal emission extends up to the TeV range, the
so-called TeV blazars.  The two main prototypes are Mrk\ 421 \citep{punch92}
and Mrk\ 501 \citep{quinn96}, two radio-loud AGN relatively close to us and
roughly at the same distance, $z_s\approx0.031$ and $z_s\approx0.034$
respectively. Five other TeV detections have been repeatedly detected (1ES\
1959+650, PKS\ 2155-304 \citep{aha05a}, 1ES\ 1426+428 and PKS 2005-489
\citep{aha05b}, and 1ES\ 2344+514 \citep{aharonian04}). All of them are
Lacertids, although it is not clear up to now whether only BL Lac objects do
emit TeV radiation or if this is due to a selection effect. As a matter of
fact, BL Lac objects appear to be much more numerous than quasars and the
closest blazars all belong to this class. A high sensitivity threshold would
strongly bias the detection toward the closest sources. Furthermore it is
well-known that TeV photons are absorbed by the Diffuse Infra Red Background
(DIrB) to create electron-positron pairs, and it is not obvious whether even
the closest quasar, 3C 273 ($z_s \approx 0.158$) could be detected in the TeV
range.

One-zone SSC models assume that highly relativistic particles are injected
in a spherical zone, where they cool by emitting synchrotron radiation and
by Inverse Compton process. The models require  specifying the source
radius, the magnetic field, as well as the density and the energy
distribution of the emitting  relativistic particles. The latter is most
often assumed to be a power law or a broken power law \citep[\emph{e.g.\
}][]{marscher83,TMG98}.  It turns out however that the computation of
emitted radiation is not compatible with the hypothesis of a static source,
because in most cases the photon density would be so high that all TeV
photons should be absorbed to form electron-positron pairs.  Furthermore
the time variability is so short (down to 15 minutes in some cases,
\citet{gaidos96}) that it is incompatible with a spherical static source
through the causality argument. This leads to assume that the source is
moving with a relativistic bulk velocity   $v=\beta c$. The effect of
relativistic bulk motion is entirely described by the Doppler beaming
factor $\delta=1/\Gamma(1-\beta\mu)$, where $\Gamma=(1-\beta^2)^{-1/2}$ is
the usual Lorentz factor and $\mu=\cos \theta$ is the cosine angle of the
jet according to the observer's line of sight. The Doppler effect shifts
all frequencies by a factor $\delta$ and all specific intensities by a
factor $\delta^3$. So the actual photon density in the jet frame is much
lower than what would be deduced for a static source. The relativistic
motion has been invoked for a long time \citep{rees66} to solve a similar
issue for radio emission of quasars.  Namely the brightness temperature  is
so high that, for a static sources, the relativistic leptons emitting
synchrotron radiation should have cooled immediately through the so-called
"Inverse Compton catastrophe" \citep{rees68}.  Again the relativistic
motion can fix this issue, because the actual photon density in the jet
frame is much lower when taking into account the Doppler amplification.
This beautiful theoretical explanation has been later confirmed by the
discovery of superluminal motion, which requires Lorentz factors at least
as great as the observed apparent reduced velocity $\beta_{\rm app}= v_{\rm
app}/c $ \citep[for a review, see\ ][]{zensus97}.\\ For $\mu \gtrsim
\beta$, corresponding to $ \theta \lesssim 1/\Gamma$, one has $1 \leqslant
\delta \leqslant 2\Gamma$, whereas $\delta \sim 1/\Gamma$ outside this
interval. It means that for a few beamed Doppler sample of boosted sources,
one expects a lot of unbeamed and not amplified counterparts. It is natural
to think that the unbeamed counterparts of bright  quasars are the weaker
radio galaxies, whose jet is thought to make a larger angle with the line
of sight.  Particularly it has been proposed that the unbeamed counterparts
of BL Lac object could be a sub class of radio galaxies, the so-called
Fanaroff-Riley I (FRI) radio galaxies  \citep{urry95}. These are
characterized by a rather faint, weakly beamed, and core-brightened radio
jet. Statistical studies of radio and X-ray AGN samples have confirmed the
possibility of such an association. The inferred beaming factors seem to
imply a rather modest value of the bulk Lorentz factor, of about 3.
However, the modeling of SSC radiation by one zone models requires much
higher values : following the authors, they range from 10 to 50
\citep{TMG98,konopelko03,sauge04a}.  The highest value seems to be needed
when one takes properly into account the extragalactic absorption.  The
problem is further complicated by the absence of clear superluminal motion
in TeV blazars, together with a rather modest brightness temperature, which
implies also a low Lorentz/Doppler factor \citep{edwards02,piner04}. All
these contradictory facts lead to what we call here the "\emph{Bulk Lorentz
factor crisis of TeV blazars}".\\

The aim of this paper is first to ascertain this crisis. We will first show
that all one-zone SSC models imply high Lorentz factors, only with the
argument of $\gamma-\gamma$ absorption and discarding any variability
argument. Then we will recall the arguments for low Lorentz factor, based
on general geometric properties of the Doppler boosting. We will show that
the explanations based on two different structures, with a possible
deceleration of a fast spine responsible for TeV emission, are not
satisfactory concerning the statistics of TeV blazars. We argue that the
best solution is to admit the low Lorentz factor constraint, abandoning the
one-zone assumption. We will show that this conclusion has important
consequences regarding the jet physics and the particle acceleration
mechanism.%%
\smallskip\\

%% --------------------------------------------------------------------------------
   \section{The case for high Lorentz factor}\label{sec:highLF}
%% --------------------------------------------------------------------------------

In the following, we will develop the need for high Lorentz factors for
one-zone models, with the fewest theoretical assumptions and relying only
on observational data. We will only assume that the SSC process is at work,
with the usual assumptions of one-zone models : the relativistic  particles
are assumed to be injected in a spherical homogeneous "blob" of radius $R$,
moving at a relativistic velocity characterized by the Lorentz factor
$\Gamma$ and a corresponding Doppler factor $\delta$. The blob is filled
with a tangled magnetic field of constant strengh $B$. We will refer to all
quantities expressed in source rest frame by a star and quantities in
observer's frame are not labeled. All energies are expressed in reduced
unit of $m_ec^2$. Throughout this paper, we express the Hubble parameter by
$H_0= 100\, h\, \rm km\,{s}^{-1}\,{Mpc}^{-1}$ and assuming $h$ to be equal
to $h = 0.65$.

%% --------------------------------------------------------------------------------
   \subsection{The synchrotron and IC differential Luminosity}
%% --------------------------------------------------------------------------------

Inspection of the TeV blazars spectra shows that the IC spectra reaches
their maximum luminosity at some peak energy $\varepsilon_c$, which is of
the order of $10^6$ for TeV  photons. This energy corresponds to an energy
$\varepsilon_c^{\ast}= \varepsilon_c \delta^{-1}$ in the blob frame. We
will consider only the particles emitting this typical energy via the IC
process, which have a typical individual Lorentz factor (in the blob frame)
$\gamma_c$, which must be greater than $\varepsilon_c^{\ast}$.\\ We then
define another typical energy $ \varepsilon_s$, that is emitted by
synchrotron process by the same particles. It can be expressed in the blob
frame as $\varepsilon_s^{\ast} = (B/B_0) \gamma_c^2$ where $B_0=3B_c/2$ and
$B_c = 2\pi m_e^2 c^3/eh \approx 4.41 \times 10^{13} \mbox{G}$ is the usual
``QED critical magnetic field strength''. One has then $\varepsilon_s =
\delta (B/B_0) \gamma_c^2$. Synchrotron spectra of TeV blazars are
typically peaking in the 1-100 keV range so that $\varepsilon_s \sim
10^{-2}-10^{-1}$.\\ 
Synchrotron photons are up-scattered at high-energy via Inverse Compton
process. It has been stressed by various authors that, giving the observed
energies of IC and synchrotron photons, the collisions between the most
energetic particles  and the peak synchrotron photons take place in the
Klein-Nishina regime, that is $\varepsilon^{\ast}_s \gamma_c \geqslant 1$. In
this condition,  the particle (electron or positron) gives all of his
energy in a single interaction. It follows that $\gamma_c \sim
\varepsilon_c^{\ast}$.  This gives an estimate of the magnetic field
strength,
\begin{equation}
\label{eq:mag}
  B = B_0 \frac{\varepsilon_s^{\ast}}{( \varepsilon_c^{\ast} )^2} = \delta B_0
  \frac{\varepsilon_s}{\varepsilon_c^2}
\end{equation}
which is only valid in the Klein-Nishina scattering regime.

The differential synchrotron luminosity $\Les$ per unit reduced energy
emitted by a population of particles of energy $\gamma$ at the energy
$\varepsilon_s$ with differential energy number of particles $\D N / \D
\gamma$ reads
\begin{equation}
\label{eq:syncr}
  \Les(\varepsilon_s) =
   \frac{\D L_s}{\D \varepsilon} (\varepsilon_s) =
\delta^3 \frac{\D L_s^{\ast}}{\D \varepsilon^{\ast}} (\varepsilon_s^{\ast}) = 
\delta^3 \frac{\D
  L_s^{\ast}}{\D N} \frac{\D N}{\D \gamma} \frac{\D
  \gamma}{\D \varepsilon^{\ast}} .
\end{equation}
The total power lost per particle of energy $\gamma$ is given by the
well-known relation $\D L_s^{\ast} / \D N = ( 4 / 3 ) c \sigma_{{\rm
Th}} \gamma^2 W_B$, and we obtain,
\begin{equation}\label{eq:lumsyn}
 \Les(\varepsilon_s) = \delta \frac{4}{3} c
  \sigma_{{\rm Th}} W_B  \frac{\varepsilon_c^3}{2 \varepsilon_s} 
  \frac{\D N}{\D \gamma}
\end{equation}
where $W_B=B^2/8\pi$ is the usual magnetic energy density. Combining with
equation (\ref{eq:mag}), we can write :
\begin{equation}\label{eq:lumsyn2}
\Les(\varepsilon_s) = \delta^3 \frac{1}{6\pi} c
  \sigma_{{\rm Th}} B_0^2  \frac{\varepsilon_s}{ \varepsilon_c} 
  \frac{\D N}{\D \gamma}.
\end{equation}
We can compute the differential IC luminosity $\Lec(\varepsilon_c)$ in the
same way using an expressionsimilar to  Eq. \ref{eq:syncr}, but replacing the
magnetic energy density by the photon energy density. However we have to
take into account that the Klein-Nishina cut-off reduces the effective
energy density available for Inverse Compton scattering. We thus define a
new characteristic energy, corresponding to the synchrotron photon energy
at the limit between the Thomson and Klein-Nishina regime for particles
with an energy $\gamma_c$. This energy is $\varepsilon_t^{\ast} =
1/\gamma_c$, i.e. in the observer's frame  
%%%%
%%%%
\begin{equation}\label{eq:epst}
\varepsilon_{t}= \delta^2/\varepsilon_{c}. 
\end{equation}
%%%%
%%%%
Photons of this energy will also be the main contributors for absorbing
photons of energy  $\varepsilon_c$ to create electron/positron pairs.If we
neglect the Klein-Nishina contribution above $\varepsilon_t$, the total
power lost per particle of energy $\gamma$ writes $\D L_c^{\ast} / \D N = (
4 / 3 ) \mbox{$c \sigma_{{\rm Th}} \gamma^2 W_{{\rm ph}}^{{\rm eff}
\ast}$}$, where
\begin{equation}
  W_{{\rm ph}}^{{\rm eff} \ast} 
  = \frac{L_s^{{\rm eff} \ast}}{4 \pi R^2 c}
  = \frac{1}{4 \pi R^2 c}
  \int_{\varepsilon^{\ast}_{\min}}^{\varepsilon^{\ast}_t } \D
  \varepsilon^{\ast} \frac{\D L_s}{\D \varepsilon^{\ast}}.
\end{equation}
For a power-law spectrum $\nu F_{\nu} \propto \nu^{\beta}$ with $\beta>0$,
a simple calculation gives
\begin{displaymath}
  W_{{\rm ph}}^{{\rm eff} \ast} \simeq \frac{1}{4 \pi R^2 c} 
  \frac{\delta^{- 2}}{\beta \varepsilon_c}\Les(\varepsilon_t).
\end{displaymath}
The coefficient $\beta$ can be replaced by another numerical coefficient
close to 1 as long as the $\nu F_{\nu}$ spectrum is growing with energy.
The differential IC luminosity reads then :
\begin{equation}\label{eq:lumic}
  \Lec( \varepsilon_c) = \delta^{- 1} \beta^{- 1} 
  \frac{\sigma_{{\rm Th}}}{3 \pi R^2}  \left. \varepsilon_c
  \Les(\varepsilon_t) \frac{\D N}{\D \gamma} \right. 
  .
\end{equation}
Comparing equation (\ref{eq:lumsyn2}) and (\ref{eq:lumic}) we can now estimate
the radius of source $R$ as a function of observed luminosities and the
unknown Doppler factor :
\begin{equation}\label{eq:radius}
  R=\delta^{- 2} \frac{3 e h}{2 \pi m_e^2 c^{7 / 2}}
  \frac{\varepsilon_c}{\sqrt{\beta \varepsilon_s}} \left( \frac {\Les(\varepsilon_t) \Les( \varepsilon_s)}{ \Lec (\varepsilon_c)}\right)^{1 / 2}
  .
\end{equation}
We will now use this radius estimate to compute the $\gamma-\gamma$ optical
depth for the photons of energy $\varepsilon_c$.

%% --------------------------------------------------------------------------------
   \subsection{The $\gamma-\gamma$ photon opacity}
%% --------------------------------------------------------------------------------

As we mention above, gamma-ray photons of energy $\varepsilon_c^{\ast}$ are
mainly absorbed by photons of energy
$\varepsilon_c^{\ast-1}=\varepsilon_t^{\ast}$ creating pairs. So the same
soft photons control both the amount of IC process and the absorption of IC
photons. The absorption probability (or opacity) per
unit path length of a photon of energy $\varepsilon_c^{\ast} \gg 1$ due to
pair production in the case of a power-law SED is given approximately by
\begin{displaymath}
  \ell_{\gamma \gamma}^{- 1} ( \varepsilon_c^{\ast} ) = \frac{\D}{\D z}
  \tau_{\gamma \gamma} ( \varepsilon^{\ast} ) = \alpha_{\gamma \gamma}
  \sigma_{{\rm Th}}  \varepsilon_t^{\ast} n (\varepsilon_t^{\ast} ) ,
\end{displaymath}
where $\ell_{\gamma \gamma} ( \varepsilon^{\ast} )$ is the free
mean path of the photon and $n ( \varepsilon^{\ast} )$ the
differential photon density per unit of reduced photon energy
$\varepsilon^{\ast}$. In the framework of one-zone model, the typical
interaction scale is of the order of the size of source $R$. It
follows that the typical $\gamma-\gamma$ optical depth writes,
\begin{displaymath}
  \tau_{\gamma \gamma} ( \varepsilon^{\ast} ) \approx \frac{R}{\ell_{\gamma
  \gamma} ( \varepsilon^{\ast} )} = \alpha_{\gamma \gamma} R
  \sigma_{{\rm Th}} \varepsilon_t^{\ast} n (\varepsilon_t^{\ast} ) .
\end{displaymath}
The function $\alpha_{\gamma \gamma}$
\citep{svensson87,coppi90} depends on the index $\beta$ of the
power-law of the spectral soft photon density expressed in $\nu
F_{\nu}\propto\nu^{\beta}$ form. A commonly used value of
$\alpha_{\gamma\gamma}(\beta)$ is $0.2$ or $0.25$. More precisely we
have \citep{svensson87}
\begin{equation}
  \alpha_{\gamma\gamma}( \beta ) 
  = 
    4^{1-\beta} 6
    \times  
    \frac{\Gamma^2 ( 2- \beta )}{\Gamma ( 7 - 2 \beta )}
    \times 
    \frac{44-\beta(41-\beta(12-\beta))}{( 4-\beta ) ( 3-\beta )}.
\end{equation}
The differential energy density number of particle is given
as a function of the differential luminosity by
\begin{equation}
  n(\varepsilon_t^{\ast}) =\frac  {\Les^{\ast}(\varepsilon_t^{\ast})}
  {4 \pi m_e c^3 R^2 \varepsilon_t^{\ast}}
  ,
\end{equation}
so we get finally the optical depth as a soft compactness at the energy $\varepsilon_t$:
\begin{equation}
  \tau_{\gamma \gamma} ( \varepsilon^{\ast} )  
  = \alpha_{\gamma\gamma}(\beta)\frac  {\sigma_{\rm Th}\Les^{\ast}(\varepsilon_t^{\ast})}
  {4 \pi m_e c^3 R} 
  = \delta^{-3}\ \alpha_{\gamma\gamma}(\beta) \frac  {\sigma_{\rm Th}\Les(\varepsilon_t)}
  {4 \pi m_e c^3 R}.
\end{equation}
Using our estimate on the source radius $R$ equation (\ref{eq:radius}), we
obtain
%%
%%
%\begin{equation}\label{eq:opa1}
  \begin{eqnarray}\label{eq:opa1}
    \tau_{\gamma \gamma} ( \varepsilon_c ) 
    &=& \delta^{- 1} 
    \tilde{\alpha}_{\gamma \gamma} ( \beta ) \frac{\sigma_{{\rm Th}} m_e
      c^{1 / 2}}{6 e h}\\\nonumber
    &&\qquad\times
    \frac{\sqrt{\varepsilon_s}}{\varepsilon_c} 
      \left( \frac{\Les(\varepsilon_t) \Lec(\varepsilon_c)} {\Les(\varepsilon_s)}
    \right)^{1/2} ,
    \end{eqnarray}
%    ,
%\end{equation}
%%
%%
where we introduce the modified function $\tilde{\alpha}_{\gamma
\gamma} (\beta)$ as $\tilde{\alpha}_{\gamma \gamma} ( \beta
)=\alpha_{\gamma \gamma} \sqrt\beta$. Values of
$\tilde{\alpha}_{\gamma \gamma}$ and $\alpha_{\gamma \gamma}$ for some
$\beta$ are tabulated in table \ref{tab:alphatau}.
%%

%% --------------------------------------------------------------------------------
   \subsection{Constraints on the local synchrotron spectral shape}
%% --------------------------------------------------------------------------------

Equation (\ref{tab:alphatau}) shows that if we are able to measure the
position in frequency and flux of both the synchrotron and the IC peak,
then we can evaluate the optical depth to $\gamma-\gamma$ absorption at the
IC peak as a function of the assumed Doppler factor value. This optical
depth is controlled by the synchrotron luminosity at the frequency
$\varepsilon_t =\delta^2/\varepsilon_c$.  We can use this relation either
by assuming some Doppler factor and evaluate the optical depth, or
constrain the value of $\delta$ by limiting the value of $\tau_{\gamma
\gamma}$. We can define $r_{\max},$ the Compton dominance parameter, as the
ratio of IC luminosity's peak to the synchrotron one, 
\begin{equation}\label{eq:rmax}
r_{\max} =\frac { 
  \left[\nu_c F_{c}(\nu_c)\right]_{\max}}
 { \left[\nu_sF_{s}(\nu_s)\right]_{\max} }= \frac {\varepsilon_c \Lec(\varepsilon_c)} {\varepsilon_s \Les(\varepsilon_s)}
.
\end{equation}
We can rewrite equation (\ref{eq:opa1}) to express the luminosity at
$\varepsilon_t = \delta^2 / \varepsilon_c $ as a function of the optical
depth and the $r_{\max}$ parameter. We finally obtain,
\begin{equation}\label{eq:cons1}
 \varepsilon_t \Les( \varepsilon_t )= \frac{\delta^4}{r_{\max} }
 \left[ \frac{\tau_{\gamma \gamma} ( \varepsilon_c^{\max}
  )}{\tilde{\alpha}_{\gamma \gamma} ( \beta )} \frac{6 e
 h}{\sigma_{{\rm Th}} m_e c^{1 / 2}}
 \frac{\varepsilon_c^{\max}}{\varepsilon_s^{\max}} \right]^2
 .
\end{equation}
Equations (\ref{eq:epst}) and (\ref{eq:cons1}) can be considered as a
system of two parametric equations of the curve giving  $\varepsilon_t
\Les( \varepsilon_t )$ as a function of $\varepsilon_t$. Eliminating $\delta$
between the two previsously cited equation, one gets the following
expression :
\begin{equation}
 \varepsilon_t \Les( \varepsilon_t )= \frac{\varepsilon_t^2 \varepsilon_c^2}{r_{\max} }
 \left[ \frac{\tau_{\gamma \gamma} ( \varepsilon_c^{\max}
  )}{\tilde{\alpha}_{\gamma \gamma} ( \beta )} \frac{6 e
 h}{\sigma_{{\rm Th}} m_e c^{1 / 2}}
 \frac{\varepsilon_c^{\max}}{\varepsilon_s^{\max}} \right]^2
 .
\end{equation}
For nearby sources, the luminosity distance writes $d_{\ell}(z) \approx c
z/H_0$ and previous expression can expressed in term on flux $F$
instead luminosity $L$ using the well known relation $F = L / 4 \pi
d_{\ell}^2$,
\begin{equation}\label{eq:cons2}
    \left( \nu_s \frac{\D F_s}{\D \nu_s}
    \right)_{\varepsilon_s \approx \delta^2 / \varepsilon_c^{\max}} 
    = 3.4 \times
    10^{-36} \delta^4\  
    \left[ \frac{h}{z}  \frac{\tau_{\gamma \gamma} (
	\varepsilon_c^{\max} )}{\tilde{\alpha}_{\gamma \gamma} ( \beta )} 
      \frac{\nu_c^{\max}}{\nu_s^{\max}}  \right]^2 r_{\max}^{- 1}
\end{equation}
For an observed SED and a given value of the opacity parameter
$\tau_{\gamma\gamma}$, the only remaining unknown quantity in the previous
equation is the beaming Doppler factor $\delta$. Each value of $\delta$
gives a point in the $\log_{10}\nu-\log_{10}(\nu F_\nu)$ plane lying on
straight line of slope 2, the level of the curve depending only on the
value of $\tau_{\gamma\gamma}$.  Intersection of the synchrotron spectrum
with the straight line directly constrains the minimum value
$\delta_{\rm min}(\tau_{\gamma\gamma})$ of the beaming Doppler factor
required to avoid the $\gamma-\gamma$ absorption with an opacity value of
$\tau_{\gamma\gamma}$ of the Inverse Compton bump (at the peak frequency).

%% --------------------------------------------------------------------------------
   \subsection{Application to Markarian 501}
%% --------------------------------------------------------------------------------

We apply this calculation to the case of the Mrk\ 501 object during the
period of the 1997 April 16 where the Beppo-SAX satellite \citep{pian98}
and the CAT imaging Atmospheric {\vC}erenkov Telescope
\citep{cat99,barrau98}  have recorded simultaneous data (see figure
\ref{fig2}).  All observational parameters we need in the equation
(\ref{eq:cons2}) are reported Table \ref{tab:parameters} .  We consider
the two cases where we take  into account, or not,  the attenuation of the
high energy component by cosmic diffuse infrared background (DIrB).  This
effect consists in the interaction of emitted gamma rays during their
travel through the Universe with the photon field of the diffuse infrared
background (DIrB) to create pairs
\citep{gould67a,gould67b,stecker92,vassiliev00}.  The tail of the high
energy spectra is then de-reddened using the method described in
\citet{sauge04a}. This situation changes the position of Inverse Compton
peak and the Compton dominance parameter $r_{\max}$.  In this case, for
$\tau_{\gamma\gamma}=1$, we obtain both in the reddened and the de-reddened
case $\delta_{\rm min}(1)\approx 50$ (see figure \ref{fig34}). 

Given equation \ref{eq:cons2}, the position of the line constraining $\delta$
depends on the value of $(\epsilon_c^{\max})^4/\Lec$. It turns out that, also
the IR un-folding of the spectrum changes both quantities, the previous ratio
depends only slightly on the level of assumed absorption. The value of
$\delta_{\rm min}(1)$ are thus quite similar in the two cases because when we
correct the Inverse Compton bump, position of the maximum moves both in
luminosity and in frequency. This effect could be clearly seen on figure
\ref{fig34}, where the difference between the two panels is hardly
perceptible.

Note that in fact the level of the curve depends implicitly also on the value
of modified power-law index of the spectrum $\beta$ (see eq.
[\ref{eq:cons2}]). In our case, we choose a value $\beta=0.5$ directly
measured on the SED. 

%% --------------------------------------------------------------------------------
   \subsection{Constraint from the variability timescale}
%% --------------------------------------------------------------------------------

Another constraint can be derived from the observation of short variability
timescale.  The classical argument is that a spherical static source cannot
be variable on a timescale smaller than $R/\delta c$. So one gets an upper
bound of radius of the source $R \leqslant R_{var,\rm min}=\delta c t_{var}
$.  Combining previous inequality to the equation (\ref{eq:radius}) and
expressing all the quantities in their fiducial units, we finally get a
 constraint similar to the one obtained in the previous section for the
local synchrotron shape (see eq. [\ref{eq:cons2}]) :
\begin{equation}\label{eq:consvar}
\left( \nu_s \frac{\D F_s}{\D \nu_s}
    \right)_{\varepsilon_s \approx \delta^2 / \varepsilon_c^{\max}} 
    \leqslant 8.3 \times
    10^{-26}\ \delta^8\  
    \left[ \frac{h}{z} \frac{\nu_s^{\max}}{(\nu_c^{\max})^2} t_{var} \right]^2 r_{\max}^{- 1}
    .
\end{equation}
Taking a characteristic variability timescale of roughly 15 min, we obtain
the left solid thick line displayed on the figure \ref{fig34}. It appears
that this constraint is less restrictive than the previous. In context of
homogeneous modeling, it gives a minimum value for Doppler factor of
$6\mbox{--}8$ and $8\mbox{--}10$ for the reddened and the de-reddened case
respectively.  

%% --------------------------------------------------------------------------------
   \section{The case for low Lorentz factor}
%% --------------------------------------------------------------------------------

In this section we shortly review all the arguments and pieces of evidence
in favour of moderate or low values of the bulk Lorentz factor.

%% --------------------------------------------------------------------------------
   \subsection{Absence of superluminal motion at parsec scale}
%% --------------------------------------------------------------------------------

Observations at the VLBI scale ($\approx$mas) show that blazars often
display superluminal apparent velocities. This phenomenon predicted by
\citet{rees66} is expected for relativistic moving sources which is highly
beamed and closely aligned with the observer's line of sight. For a
component moving along the jet axis at a reduced speed $\beta = v/c$ and
making an angle $\theta$ from the line of sight, the apparent transverse
velocity measured by the observer is :
\begin{equation}\label{eq:slvelocity}
  \beta_{\rm app}=
  \frac{ \beta\sin\theta }{ 1-\beta\cos\theta }
  \leqslant \beta\Gamma.
\end{equation}
If $\beta>\beta_{\rm crit}=\sqrt{2}/2$ and $\theta$ is such that $\sin
2\theta>(\Gamma^2-1)^{-1}$, the motion will appear to be superluminal
{\itshape i.e.} $\beta_{\rm app}>1$. Expressed in degrees the latter
condition writes $\theta>\theta_{\rm crit}=0.28\, (\Gamma/10)^{-2}\ \rm
deg$.  

As a matter of fact,  VLBI/VLBA campaigns have not clearly succeeded in
finding superluminal motion at the parsec scale for any TeV blazars
\citep{edwards02,piner04}.  Observed radio components seem to be stationary
or subluminal, requiring low or moderate values of the Lorentz factor
($\Gamma\approx2-4$).\\
The absence of superluminal motion could be explained by a very close
alignment of the jet with the line of sight. Indeed, following the previous
expressions, if $\theta\leqslant 1/2\Gamma^2$ the apparent velocity is
always smaller than $c$, and object appears to be subluminal despite the
large value of $\Gamma$. But it this case, a simple statistical argument
based on the density number of unbeamed counterparts rule out this
possibility as we will see in the next section.\\

Moreover, derived value of the brightness temperature of the VLBI core is in
the order of $10^{10-11}{\ \rm K }$ and lie well below the usual Inverse
Compton limit of $\approx 10^{11-12}\ \rm K$ necessary to avoid the "Inverse
Compton catastrophe", {\itshape i.e.} situation where ultra-relativistic
particles suffer from dramatically Compton cooling in a very short time.
\citet{piner04} have concluded that the jet should be only mildly relativistic
at parsec scale. They propose that the TeV emitting inner jet is strongly
decelerated before reaching the parsec scale.  However we will see in the
following that the existence of the highly relativistic motion is challenged
by other observational facts concerning the statistics of beamed \textit{vs.}
unbeamed sources.

%% --------------------------------------------------------------------------------
   \subsection{%%
        Number of beamed sources\\
        in the BL Lac/FR-I unification paradigm}\label{sec:statobj}
%% --------------------------------------------------------------------------------

As we said in the introduction, the blazar phenomenon arises from a close
alignment of jet axis with the observer's line of sight. Following this
scheme, one expects the existence of sources sharing the same physical
properties (\emph{i.e.} intrinsically the same objects), but viewed at
larger angle. It has been proposed that Fanaroff-Riley radio galaxies can
be the unbeamed parent population of blazars and particularly, FR-I
galaxies can be the counterparts of Lacertids \citep{urry95}.  \\ The
unification hypothesis can be tested on samples of objects both by their
luminosity ratio and by their spatial density.  Doppler beaming effect
enhances the intrinsic bolometric luminosity by a factor $\delta^4$.  The
Doppler factor itself varies from $\delta_{\rm max} = 2\Gamma$  for a jet
pointing exactly toward the observer, to $\delta_{\rm min} = 1/\Gamma$ for
jets lying close to the sky plane. Although the exact definition of what is
a beamed object can be somewhat subtle, one can estimate an order of
magnitude of the number of such objects. It is easy to see that the solid
angle for which the Doppler factor is larger than some given value
$\delta_0$ (where of course $\Gamma^{-1} <\delta_0< 2\Gamma$ is
\begin{equation}\label{eq:omega}
\Omega = 2 \pi (1-\mu_0) \simeq \frac {2 \pi }{\Gamma} \left( \frac{1}{ \delta_0 } - \frac{1}{ 2 \Gamma }
\right)
\end{equation}
where we used $1/\beta \simeq 1 + 1/(2 \Gamma^{2})$, and hence the fraction
of sources with a Doppler boosting larger than $\delta_0$ is approximately
\begin{equation}
f(\delta>\delta_0) \simeq \frac {1 }{\Gamma} \left( \frac{1}{ \delta_0 } - \frac{1}{ 2 \Gamma }
\right)
,
\end{equation}
if we assume always two symmetrical jets.\\ 
So any "beaming criterion" imposing a Doppler factor larger than some
sizable fraction of $\Gamma$ will give a fraction of beamed sources of the
order of $\Gamma^{-2}$. For one beamed source, one expects thus around
$\Gamma^2$ unbeamed sources.  It turns out that careful statistical studies
do indeed confirm the association of BL Lac objects with FR-I galaxies, but
they converge toward a much lower Lorentz factor than what is expected from
the gamma-ray emission. For X-ray selected BL Lacs, which comprise all
known TeV blazars, the inferred density ratio is 1:7, corresponding to a
bulk Lorentz factor around 3.5  \citep{urry95}. 

In the same way, we can examine the hypothesis that the lack of detection
of superluminal motion would be due to a close alignment of the jet with
the line of sight.  As we note in the previous section, a beamed source
with $\theta\leqslant 1/2\Gamma^2$ can not appear superluminal. The cone
substained by this angle corresponds to the solid angle $\Omega\approx
\pi/4\Gamma^4$.  The ratio $f'$ between the density of  unbeamed sources
and subluminal beamed sources one is thus given by
\begin{equation}
        f'
        =\frac{n_{\rm FR-I}}{n_{\rm TeV,sub}}
        \approx16\Gamma^4.
\end{equation}
Observations show that 5 TeV blazars do not clearly display superluminal
motions for $z_s\leqslant 0.047$ \citep{edwards02,piner04}. Assuming
$\delta = 2\Gamma = 50$, it corresponds roughly to a volume of $0.043\rm\
Gpc^3$ and then to a density of subluminal TeV blazars of $n_{\rm
TeV,sub}=117\rm\ Gpc^{-3}$.  Then, the density of expected unbeamed
counterparts would be $n_{\rm FR-I}\approx 7.3 \times 10^{8}\rm\ Gpc^{-3}$
which is absolutely unreasonable.

%% --------------------------------------------------------------------------------
   \subsection{%%
        Luminosity ratio of beamed sources\\
        in the BL Lac/FR-I unification paradigm }
%% --------------------------------------------------------------------------------

Another constraint can be derived from the luminosity ratio between the
Lacertids sources and the FR-I ones. Supposing the same assumption as above
(see previous section \ref{sec:statobj}), \textit{ie} Lacertids and FR-I
are on average the same intrinsic objects but viewed at different angles,
the bolometric luminosity contrast between the two parents population
(Lacertids and FR-I radio galaxies) is given by,
\begin{equation}\label{eq:ratbol1}
        \varpi
        = \frac{\mathcal{L}_{\rm Lac}}{\mathcal{L}_{\rm FRI}}
        = \left(\frac{\delta_{\rm Lac}}{\delta_{\rm FRI}}\right)^4.
\end{equation}
In the case of Lacertids, relativistic beaming requires
$0\leqslant\theta\leqslant 1/\Gamma$ or equivalently
$2\Gamma\geqslant\delta_{\rm Lac}\geqslant \Gamma$.  On the other hand, we
suppose that off-axis counterparts verify $\delta\approx2/\Gamma$
(corresponding to an average angle value of $\theta\approx60\,\deg$ for
$\Gamma>1$). Then, equation (\ref{eq:ratbol1}) rewrites
\begin{equation}\label{eq:ratbol2}
    \Gamma^8
    \geqslant \varpi \geqslant
    \frac{\Gamma^8}{16}.
\end{equation}
This estimate can of course be complicated by an intrinsic luminosity
distribution.  It may be also that we cannot detect the unbeamed sources
due to limited sensitivity of the instrument. However some other indicators
such as the extended radio lobes power or the galaxy luminosity itself are
not highly beamed, and can serve as an unbiased criterion to select
samples. \\ 

\citet{capetti99} studied a sample of 12 Lacertids and 5 FR-I sources with
HST and compared the core luminosity ratio between objects sharing similar
radiative properties. It clearly appears that the whole emission of the
Lacertid cores is roughly $10^2$--$10^5$ times brighter than the
corresponding radio galaxy ones. Moreover \citet{chiaberge00} performed a
similar work on a larger and more complete sample. They roughly obtained
the same conclusions : the luminosity ratio between Lacertids and radio
galaxy belongs to the interval $10^{2.5}$--$10^{5.5}$. Applying relation
(\ref{eq:ratbol2}), we obtain typical values of $\Gamma\approx2\mbox{--}5$
for the bulk Lorentz factor.  They also compare the broad band spectra of
both classes of objects and they found that the spectra could be deduced by
a simple Doppler boosting, but once again with modest values of the Doppler
factor. 

%% --------------------------------------------------------------------------------
   \subsection{Detection of TeV unbeamed source -- the case of M87}
%% --------------------------------------------------------------------------------

The nearby giant elliptical radio galaxy M87 (NGC4486, $z_s\approx
0.00436$) is the first (and for the time being unique) detected unbeamed
radio-loud source at the TeV energy range. First detection was reported by
the HEGRA collaboration with an integral flux above 250 GeV at about
$3.3\%$ of the flux of the Crab Nebula (with a significance of
$4.7\,\sigma$) during an high state \citep{aharonian03,beilicke04}. Such
TeV events are confirmed by recent measurements of \textit{High Energy
Stereoscopic System} (HESS) \citep{beilicke05}.
The powerful radio jet of M87 has been well studied in various wavelengthes
from radio to X-rays, showing that the jet axis makes an angle between 30
and 40$\,\deg$ with the line of sight. This angle is clearly large enough
to ensure that the emission is unbeamed. Previous works based on the study
of the proper motion of the VLBI knots \citep{biretta95} or on the detailed
analysis of HST and VLA observations \citep{lobanov03} converge toward
value for  $\Gamma$ of 3--5 at the kiloparsec scale. 
The jet differential flux of a source expressed in the observer's frame can
be written  as function of the intrinsic differential luminosity as
\begin{equation}
        \mathcal{F}_{\nu}(\nu;\theta,z)
        \approx 
        (1+z)\delta^3\  
        \frac  {\mathcal{L}_{\nu}^{*} (\nu^{*})} {4 \pi d_{\ell}^2}
        ,
\end{equation}
with $\nu=\nu^{*}\delta/(1+z)$ and where  $d_{\ell}(z)\approx zc/H_0$ is the
usual luminosity distance.  We now consider two
different versions of the same intrinsic object, a beamed one corresponding
to a blazar and an unbeamed one corresponding to a radio galaxy.
In this case, the ratio $\mathcal R$ of the observed photon fluxes above some
threshold frequency $\nu^{\rm thr}$ writes,
    \begin{equation}
    \begin{array}{rcl}
        \mathcal{R}=
        \displaystyle
        \frac
          {\int_{\nu^{\rm thr}} {\rm d}\nu\, [{\mathcal F_{\nu}^u(\nu)}/{h\nu}]}
          {\int_{\nu^{\rm thr}} {\rm d}\nu\, [{\mathcal F_{\nu}^b(\nu)}/{h\nu}]}
        &=&
        \displaystyle
        \left(\frac{z_b}{z_u}\right)^2
        \left(\frac{1+z_b}{1+z_u}\right)^{\alpha-2}
        \left(\frac{\delta_u}{\delta_b}\right)^{2+\alpha},\\
        &=&
        \displaystyle
        k(z_b,z_u;\alpha)
        \left(\frac{\delta_u}{\delta_b}\right)^{2+\alpha},
    \end{array}
    \end{equation}
where the index ${}_u$ (resp. ${}_b$) refers to the observed unbeamed
(resp. beamed) quantities, and where we suppose that the high energy
spectrum can be expressed as a simple power-law with a photon index
$\alpha$.

For beamed sources, the Doppler factor can be written as $\delta_b\approx
2\Gamma$ while for the unbeamed case one has
$\delta_u=1/\Gamma(1-\beta\cos\theta)<\Gamma$ with $\beta\approx
1-1/2\Gamma^2$. Finally we can express the bulk Lorentz factor as a function
of   $\theta$ and the observational
parameters only
    \begin{equation}
        \Gamma(\theta)
        =
        \left\{
        \frac
        { \left[k(z_b,z_u;\alpha)\mathcal{R}^{-1}\right]^{1/(2+\alpha)}-\cos\theta }
        { 2(1-\cos\theta) }
        \right\}^{1/2}.
    \end{equation}
A raw approximation of the previous expression is 
    \begin{equation}\label{eq:crude}
        \Gamma(\theta)
        \approx
        \frac{1}{\theta}\,
        \left[\frac{k(z_b,z_u;\alpha)}{\mathcal{R}}\right]^{1/2(2+\alpha)}
        ,
    \end{equation}
showing the $\sim 1/\theta$ functional dependence of $\Gamma$ and its slow
power-law variation with $\mathcal R$ (or $k$). For instance, for a typical
value of $\alpha=2.5$, a factor ten on $\mathcal R$ implies  only a factor
($10^{1/9}\approx 1.29$) on $\Gamma$. 

In 1997 April flaring period, the TeV blazar Mrk~501 ($z_s\approx 0.034$)
became roughly 8 times as bright as the Crab Nebula as reported by the
French collaboration CAT \citep{cat99}. Assuming M87 is an unbeamed
counterpart of Mrk~501 with an angle $30 \deg\leqslant\theta_{\rm
M87}\leqslant40\deg$ we obtain $4\leqslant\Gamma\leqslant 5.3$. 
Again we find that the luminosity ratio is compatible with modest values of
the Lorentz factor.
Due to the increasing sensitivity of the present and the next
generation of the Imaging Atmospheric {\vC}erenkov Telescope Arrays, the
detections of more and more TeV radio galaxies should help us to constrain
the dynamics of the emitting plasma at the subparsec scale in a more
reliable statistical way. 

%% --------------------------------------------------------------------------------
   \subsection{Summary}
%% --------------------------------------------------------------------------------

All the above considerations show that observational data are compatible
with the beaming model only if the bulk Lorentz factor for the X-ray and
TeV emitting part of the object is relatively low, between 3 and 5. This
value reproduces correctly the luminosity ratio \emph{and} the statistical
number of sources (which are a priori independant factors).  Conversely, a
value of $\Gamma=12.5$ which is the minimum typical value derived from the
one-zone modeling approach, would lead to a luminosity contrast of $
\varpi\approx 10^{7.6}\mbox{--}10^{9} $. This latter estimation is
clearly not compatible with the previous observations, ascertaining the
"\emph{Bulk Lorentz factor crisis of TeV blazars}". In the following, we
will examine some suggestions made by various authors to solve the crisis.

%% --------------------------------------------------------------------------------
   \section{How to solve the crisis}
%% --------------------------------------------------------------------------------

%% --------------------------------------------------------------------------------
   \subsection {Two pattern model}
%% --------------------------------------------------------------------------------

\cite{chiaberge00} and \cite{trussoni03} argue that a jet velocity
structure can solve the problem of the BL Lac/FR-I unification scheme.
They consider a (med-)relativistic external layer and a fast internal spine
which dominates the emission in the case of a favorable alignment along the
observer's line of sight, \emph{i.e.}  in the blazar case. Although similar
in appearance to the two-flow model of \citet{pel85} (\textit{see below for
details}), it differs by the fact that both flows are relativistic, one
with a "low" Lorentz factor (around 3) and one with a high Lorentz factor
(at least 10).  In the following, we consider the same approach considering
a two-components modeling of the velocity structure, where a fast inner
structure is supposed to be surrounded  by a slow one.  Each of these
components is respectively characterized by a bulk Lorentz factor
$\Gamma_{\rm f}$ and $\Gamma_{\rm s}$. As we saw, the radiative emission of
the moving source with a bulk Lorentz factor $\Gamma$ is beamed in a cone
sustained by a solid angle $\delta\Omega=\pi/\Gamma^2$ along the motion.
Therefore the emitted radiation appears to be Doppler boosted when the jet
lies into $\delta\Omega$ around the observer's line of sight. In this case,
the luminosity contrast between the two parents population
(Lacertids and FR-I radio galaxies) writes,
\begin{equation}\label{eq:ratio2}
	\Gamma_{\rm s}^8 
 	\leqslant
	\varpi
	\leqslant
 	(\Gamma_{\rm s}\Gamma_{\rm f})^4,
\end{equation} 
where right and left bound correspond to the case where fast velocity
component respectively dominates or not the emission. Unification models
are sensitive to the slow component only, so ($\Gamma_{\rm s}=3-5$)
\citep{hardcastle03,trussoni03,chiaberge00,urry91}. Therefore, assuming
$\Gamma_{\rm s}\approx 4$ we obtain $10^{4.81}\leqslant\varpi\leqslant
10^{8}$ if $\Gamma_{\rm f}=25$ and $10^{4.81}\leqslant\varpi\leqslant
10^{6.4}$ if $\Gamma_{\rm f}=10$.\\
The results give a possible solution to the luminosity ratio problem, being
more compatible with observations. But in the next section we will examine
the consequences of such a velocity structure on the detection probability
of TeV emitters among BL Lac objects.

%% --------------------------------------------------------------------------------
   \subsubsection{Statistics of detected sources}
%% --------------------------------------------------------------------------------

Suppose a population $\Sigma$ of sources randomly oriented per unit
volume $n_0$. Then the density number of sources oriented with an angle
$\theta = \cos^{- 1} \mu$ according to the observer's line of sight is
${\D n}/{\D \mu}={n_0}/{2}$. We define BL Lac sources as those seen into the
$\delta\Omega_{\rm s}=\pi/\Gamma^2_{\rm s}$ cone and therefore the TeV
emitters as the part of Lacertids lying into the $\delta\Omega_{\rm
f}=\pi/\Gamma^2_{\rm f}$. Therefore the probability of detecting a
BL Lac object in $\Sigma$ writes
\begin{equation}
 \mathcal P_{\rm Lac}=\mathcal P(\mu\leqslant\mu_0)
 =\frac 1{n_0}
  \int_1^{\mu_0}\D\mu\, \frac{\D n}{\D\mu}=\frac 1{4\Gamma^2_{\rm s}},
\end{equation}
where $\mu_0=1-1/2\Gamma_{\rm s}$. The probability that a BL Lac object
$\omega\in\Sigma$ is also a TeV emitter source (\textit{i.e.} a source
which the emission is dominated by the fast inner structure) is given by
the conditional probability,
\begin{equation}\label{eq:probindiv}
 \mathcal P_{\rm TeV/Lac}
 =\mathcal P({\omega\in{\rm TeV}|\omega\in{\rm Lac}})
 =\left(
 \frac{ \Gamma_{\rm s} }{ \Gamma_{\rm f} }
 \right)^2.
\end{equation}
The probability of detecting $n$ TeV emitters among a population of $N$
Lacertids is given by the usual binomial probability law,
\begin{equation}\label{eq:prob1}
\mathcal P(n/N)=\frac{N!}{n!(N-n)!}\,
\mathcal P_{\rm TeV/Lac}^n(1-\mathcal P_{\rm TeV/Lac})^{N-n}.
\end{equation}
and implicitly depends on the value of the ratio $\Gs/\Gf$. 
It is more convenient to express the probability of detection of at least
$n$ TeV emitters among the same sample of $N$ Lacertids which writes,  
\begin{equation}\label{eq:cumprob}
	\mathcal P(N\geqslant n_0\geqslant n)
	=\sum_{k=n}^{k=N}\mathcal P(k/N)
        .
\end{equation}

%% --------------------------------------------------------------------------------
   \subsubsection{Applications}
%% --------------------------------------------------------------------------------

For a given value of $(n,N)$, requiring that $\mathcal P(N\geqslant
n_0\geqslant n)$ is larger than an a priori probability $\mathcal{P}_0$
constrains the space parameters $(\Gs,\Gf)$. The latter inequality leads to
eliminate parameter's region lying above a straight line which  corresponds
equivalently to a constant value of $\mathcal P_0$ or of the ratio
$\Gs/\Gf$ (see eq.  [\ref{eq:cumprob}], [\ref{eq:prob1}] and
[\ref{eq:probindiv}]). Further restriction come from the $\gamma-$rays
transparency argument and FR-I/Lacertids unification models as developed
above.
\begin{enumerate}
    \item Firstly, the $\gamma-$rays transparency argument developed in the
    first part of this work directly constrains the value of the fast
    component as it requires a minimum value of the Doppler factor
    $\delta_{\min}$, and therefore $\Gf\geqslant \delta_{\min}/2$. We have
    shown that $\delta_{\min}\approx 50$ exclude all part of the
    $(\Gs,\Gf)$ lying above $\Gf=25$.
    \item Secondly, basic statistics argument based on the number FR-I
    radio galaxies regarding the Lacertid one and the comparison of
    luminosity distribution of the previous populations constrain the value
    of the slow component to reasonable values less than $\Gs\approx 7$
    \citep{urry95,chiaberge00}.
\end{enumerate}
We test this result on the catalog of BL Lac objects from \citet{giommi95}.
At $z_s\leqslant 0.13$ they report 29 Lacertids with known redshift.
Setting $(n=7,N=29)$, $\mathcal{P}_0=1\%$ and recalling that
$\Gs{}_{,\max}=7$ and $\Gf{}_{,\min}=25$, the intersection of all listed
previous constraints reduces to null region (see figure \ref{fig6}). Even with 
  the hypothesis of a structured flow, a large value of the Lorentz
factor recquired by one-zone homogeneous models is clearly untenable
(excluded with a confidence level of $99\%$).

We demonstrate that even if the two-components velocity structure can give
a satisfactory answer to the luminosity problem of the Lacertids even with large
value of Doppler factor required by high energy emission models, it fails
to explain the detection statistics of the TeV emitters among the BL Lac
object population supposed to be off-axis FR-I sources. 

%% --------------------------------------------------------------------------------
   \section{Discussion}
%% --------------------------------------------------------------------------------

%% --------------------------------------------------------------------------------
   \subsection {Inhomogeneous models}
%% --------------------------------------------------------------------------------

Altogether, the previous considerations lead to a serious paradox, where a
high Lorentz factor larger than 20 seems mandatory to avoid strong
$\gamma-\gamma$ absorption, whereas all other facts tend to favour modest
values around 3. The only way to solve the discrepancy seems to give up the
implicit assumption of all one zone model, i.e. the fact that all photons
are produced co-spatially and simultaneously in some characteristic region
of size $R$. Alternative to one-zone models have already been discussed in
the literature. For instance, in the "blob-in-jet" model
\citep{sol01,sol03}, low energy photons are produced in a continuous jet
and only the high energy ones are produced in a spherical blob. This allow
to fit the overall spectrum with a smaller Doppler factor of around 15.
Another possibility is to take explicitly the variability and use a
time-dependant model to reproduce the data. Again, the constraints arising
from $\gamma-\gamma$ opacity can be somewhat released because soft photons
are emitted at a later stage than high energy ones. As has been remarked by
\citet{ghisellini85}, time-dependant model will produce effects comparable
to inhomogeneous ones. If the evolving source is moving at relativistic
velocity, and that many flares are contributing to the emission, the
overall system will be in fact an stratified jet composed with many
"one-zone" regions in a different evolutionary stage. However, none of
these models do use bulk Lorentz factors as low as 3. \\

We are led thus to consider models where photons are distributed along a
jet in a continuous structure, instead of filling a spherical source. In
this case, the luminosity is proportional to the photon density times the
lateral surface of the jet, which is  $2 \pi R_j h_j = 2 \pi A R_j^2$,
where $R_j$ and $h_j$ are the typical  jet radius and length at the
emission region, and $A = h_j/R_j$ is an aspect ratio of the source. For a
self similar jet for which all quantities (radius, magnetic field, particle
density etc...) are described by power law as a function of the distance
$z$, one expects $h_j \sim z_j$, where $z_j$ is the distance of the
emitting region from the center. It follows that $A  \sim  z_j/R_j \sim
\theta_j ^{-1}$, where $\theta_j$ is the typical opening angle of the jet.
One can see that for a given synchrotron luminosity and photon density
(implying the same IC luminosity), one must conserve the quantity $A
R_j^2$, so the typical radius of the jet, and hence the $\gamma-\gamma$
optical depth will be reduced by a factor $A^{1/2}$ with respect to a
spherical source. This simple geometrical modification helps thus to
increase luminosity without increasing optical depth.  Furthermore, the
particle distribution needs not to be the same all along the jet. Rather
one expects a gradual  cooling of the particles, the overall spectrum being
the envelope of all slices of the jet. The local photon spectrum can thus
be different from the observed one, and particularly the local soft photon
density can be much lower, helping again to reduce the $\gamma-\gamma$
optical depth. As we shall see, all these factors can offer a clue to the
Bulk Lorentz factor crisis, but imply strong constraints on the physical
picture of relativistic jets.

%% --------------------------------------------------------------------------------
   \subsection {Theoretical implications}
%% --------------------------------------------------------------------------------

%% --------------------------------------------------------------------------------
   \subsubsection{Local photon density}
%% --------------------------------------------------------------------------------

Considering the above constraints, we will take the opposite attitude,
considering that the value of the bulk Lorentz factor is constrained by the
unification models and the detection of unbeamed sources to be around 3.
The typical high energy emission zone as defined above is equivalent to the
superposition of $A$ spherical sources with individual luminosities
$L_{\nu} / A \sim  \theta_j L_{\nu}$.  Therefore, all previous equations in
section \ref{sec:highLF} are still valid provided we replace the observed
luminosity $L_{\nu}$ by $\theta_j L_{\nu}$. Using equation (\ref{eq:opa1}),
we conclude that all opacity constraints remain unchanged  if we replace
the optical depth $\tau_{\gamma\gamma}$ by
$\theta_j^{-1/2}\tau_{\gamma\gamma}$, which is of course in accordance with
the estimate made in the previous paragraph. So we can use the figure
\ref{fig34} with slightly different values of $\tau_{\gamma\gamma}$. The
typical angle $\theta_j$ must be of the  order of $10^{-2} - 10^{-1}$, so
the optical depth will be reduced by a factor between 3 and 10. In the
following, we will still use the same line
$\theta_j^{-1/2}\tau_{\gamma\gamma} \approx 1 $ to constrain the optical
depth, meaning that $\tau_{\gamma\gamma} \leqslant 0.1$ to $0.3 $.\\

%% --------------------------------------------------------------------------------
   \subsubsection{Local photon spectrum}
%% --------------------------------------------------------------------------------

We can thus put an upper limit on the soft photon luminosity corresponding
to this value and a Doppler factor of 3, which constrains the soft photon
luminosity at an energy $\varepsilon_t =\delta^2/\varepsilon_c \sim 10\
{\rm eV}$. As the spectrum is by definition approximately the same in all
the characteristic emission region, we can thus estimate the local photon
spectrum by interpolating between the peak synchrotron luminosity and the
above upper limit. Inspection of figure \ref{fig34} shows that the spectral
index between 10 eV and 100 keV is very close to 1/3, which is
characteristic of a quasi monoenergetic distribution; an example of such
distribution is provided by the quasi-maxwellian or "pileup" distribution
\citep{hp91,schlickeiser85,sauge04a}, which is a natural outcome of some
acceleration  processes like second order Fermi acceleration or magnetic
reconnexion. This distribution is not the usual power law often claimed to
exist in AGN, and which is naturally produced in MHD shocks. Rather than
localized shocks, the assumption of low Lorentz factor leads to a picture
of a continuous jet filled by relativistic particles, continuously reheated
by a diffuse acceleration mechanism. 

%% --------------------------------------------------------------------------------
   \subsubsection{Pair production}
%% --------------------------------------------------------------------------------

The local synchrotron spectrum can not be harder than the monoenergetic
one; so figure \ref{fig34} proves that this implies a lower limit to the
quantity $  \theta_j^{-1/2}  \tau_{\gamma\gamma}  \gtrsim1 $. Thus the
limit on $\gamma-\gamma$ optical depth can not be very low, unless we have
an extremely well collimated jet that is not supported by the general FR-I
morphology. Modest collimation factors imply that  $\tau_{\gamma\gamma}
\gtrsim 0.1$. This supports the formation of an electron-positron pair
plasma in the acceleration site. If the acceleration is not localized,
which is suggested by the picture of a continuous jet filled by a pile-up
distribution, the pairs created by ${\gamma-\gamma}$ interaction can not
avoid being reaccelerated and will trigger a pair cascade \citep{hp91}.
These constraints are thus suggestive of a inner continuous pair dominated,
jet-like emission zone, maintained at a relativistic temperature. 
 
%% --------------------------------------------------------------------------------
   \subsection{Compatibility with the two-flow model}
%% --------------------------------------------------------------------------------

All the previous considerations find a natural explanation in the context of
the two-flow model, which was proposed to account for the formation of
relativistic jets in AGN : in this model, extragalactic jets are in fact
the results of a double structure: a first jet, not highly but only mildly
relativistic ($v \approx 0.5 c$), is emitted by a MHD mechanism by a large
scale magnetic field anchored in an accretion disk
\citep{bp82,fer93a,fer93b,fer95} ; this powerful, but weekly dissipative
jet, can sustain a MHD turbulence able to accelerate non thermal particles.
These particles will produce synchrotron and gamma-rays photons, and if the
optical depth becomes large enough, these photons will trigger an intense
pair cascade leading to a dense pair plasma in the empty "throat" of the
jet. We have shown in previous works that this pair plasma will be
spontaneously accelerated to relativistic velocities even if the
surrounding jet is not highly relativistic by itself, by the so called
"Compton Rocket effect", which is a recoil effect associated with
anisotropic IC process originally introduced by \citet{odell81}. The
Compton Rocket effect has been shown to be inefficient to accelerate an
\emph{isolated} relativistic plasma because the cooling time is always
shorter than the bulk acceleration time \citep{phinney82}. In the two-flow
model however, the heating by the surrounding jets compensates for the
cooling and the pair plasma remains relativistic over large distances
\citep{marcowith95}. \\

Detailed calculations of the Compton Rocket effect in this configuration
\citep{renaud98} show that the pair plasma accelerates gradually in the
vicinity of an accretion disk, being maintained to a quasi equilibrium
Lorentz factor $\Geq\approx (z/r_i)^{1/4}$, where $r_i$ is the inner radius
of the accretion disk (3 gravitational radii for a Schwarschild black
hole).  The equilibrium Lorentz factor is defined by the fact that the
photon field of the accretion disk, \emph{seen in the comoving frame},
appears to be nearly isotropic due to relativistic aberration. It grows
slowly with the distance,because the field becomes more and more
anisotropic.  The acceleration continues until the photon density becomes
to low to efficiently accelerate the plasma. Then the plasma decouples from
the ambient radiation field and ends up with an asymptotic balistic motion
at constant $\Gamma_{\rm b}\to\Ginf$, which depends on the disk luminosity
and the particle energy distribution.  For a relativistic energy
distribution function $n(\gamma) \propto \gamma^2 \exp
(-\gamma/\bar{\gamma})$, the asymptotic bulk Lorentz factor is
approximately $\Ginf\approx(\ell_s \bar{\gamma})^{1/7 } $ , where $ \ell_s
= L_s \sigma_{\rm Th}/ 4\pi m_e c^3 r_i $ is the soft photon disk
compactness and $\bar\gamma$ is the characteristic energy of the pileup
depending on the details of the acceleration/cooling processes
\citep{renaud98}.\\

The first interesting feature in this model is that it predicts naturally a
gradual acceleration from the core. The value of $\Gamma_{\rm b} \approx 3$
is naturally obtained at $ \approx 100\ r_g$, which is a typical distance
where gamma-ray emission seems to occur, based on variability arguments.
Thus low Lorentz factor are not surprising in this model, but are explained
naturally. As a matter of fact, very high values of 20 near the core  would
be difficult to explain in this frame!\\

The second one is that the asymptotic bulk Lorentz factor is controlled by
the density of the photon field emitted by the accretion disk. For BL Lac
objects and FR-I galaxies, the disk luminosity is known to be much lower
than luminous FSRQ and FR-II galaxies, by a factor around $10^{-3}$.  One
would expect thus a lower asymptotic Lorentz factor for BL Lac object in
average, which would help to understand the absence of superluminal motion
in TeV blazars. As a matter of fact, numerical estimates show that the
expected asymptotic Lorentz factors are between 10 and 20 for
near-Eddington accreting supermassive black holes, whereas they are rather
between 5 and 10 for low luminosity AGN. We note that bulk Lorentz factors
around 5 are indeed observed in M87, which would mean that the decoupling
occurs at some thousands Schwarzschild radii from the core. Unification
models are compatible with slowly accelerating jets, the inner (X-ray
emitting) jets having bulk Lorentz factors around 3 and the outer radio jet
having a larger Lorentz factor around 7 (\cite{urry95}). Again this is
perfectly compatible with the predictions of the two-flow model, with an
inner jet emitting X-ray and TeV radiation with a modest bulk Lorentz
factor, and an outer jet responsible for radio emission with a higher one.
Also we note that there is no need for deceleration  to explain FR I mildly
relativistic jets : even if the large scale jet has only a moderately
relativistic velocity $v\approx 0.5 c$, this can be attributed to the
"slow" MHD component surrounding the relativistic beam, the latter being
dissipated at kpc scale. 

An inhomogeneous model offers also a convenient explanation for the lack of
obvious correlation between X-rays and gamma-rays variability. If the magnetic
field is varying along the jet, the photons with a given energy could be
produced by electrons with different energies and locations. If several flares
contributes to the observed spectrum --- which is necessary to account for the
global spectral shape in case of a monoenergetic distribution --- a
complicated variability pattern could emerge. This is much less easy to
understand in the homogeneous steady state models. Thus we think that
inhomogeneous models, although more complicated to compute, seem to be
unavoidable to explain the spectral and temporal features of TeV blazars'
emission. 

%% --------------------------------------------------------------------------------
   \section{Conclusion}
%% --------------------------------------------------------------------------------

We have investigated in detail the so-called "Bulk Lorentz factor" crisis
of TeV blazars, which seem to imply an incompatibility between a high
Lorentz factor required to insure gamma-ray transparency, and a low Lorentz
factor deduced from statistical arguments and luminosity contrast,
including the detection of the non-blazar TeV source M87. We show that the
transparency argument is common to all one-zone models, and that the only
way of solving the paradox is to consider inhomogeneous jet models, where
all photons are not produced cospatially. The spectrum is then the spatial
convolution of different jet slices, and the opacity problem can be avoided
by invoking geometrical arguments and harder local photon spectrum. We show
however that for modest values of geometrical beaming of the jet, which
seem natural considering the morphology of FR-I galaxies, the optical depth
for $\gamma-\gamma$ absorption can not be very low, even for a local
quasi-monoenergetic particle distribution. This has profound implications
on the physics of the jet : the acceleration mechanism must be distributed
all along the jet, and is more probably insured by second order Fermi
mechanism or reconnexion sites than by localized shocks . A moderately high
value of  $\gamma-\gamma$ optical depth implies a fair production rate of
electron-positron pairs, which are likely to be reaccelerated by the
acceleration  process to trigger a pair cascade. All this features are
natural consequences of the two-flow model, which attributes the
relativistic phenomena (high energy emission and superluminal motion) to
the formation of such a pair plasma inside a powerful, but mildly
relativistic jet insuring the confinement and the heating of the
relativistic beam.  The bulk Lorentz factor is also well in accordance with
a continuous acceleration along the jet by the Compton Rocket effect, which
predicts naturally $\Gamma_{\rm b} \approx 3 $ at a hundred Schwarzschild
radii from the core. We conclude that all observational facts are more in
accordance with light, moderately relativistic leptonic beams than with
highly relativistic baryonic jets. 
%% 
%%%%%%%%%%%%%%%%%%%%%%%%%%%%%%%%%%%%%%%%%%%%%%%%%%%%%%%%%%%%%%%%%%%%%%
%% 
\acknowledgments{%
Remarks of an anonymous referee helped to improve significantly the final
version of this paper. L.S. would like to thank all members of the IPNL team
of the SNFactory collaboration. Part of the simulations reported here has been
performed at the ``\emph{Centre de Calcul Intensif de l'Observatoire de
Grenoble.}''.
}
%% 
%%%%%%%%%%%%%%%%%%%%%%%%%%%%%%%%%%%%%%%%%%%%%%%%%%%%%%%%%%%%%%%%%%%%%%
%%

%%
%%
\clearpage
\begin{deluxetable}{lcccccc}
\tablecaption{Mkn 501 1997 April-16 observationnals parameters\label{tab:parameters}}
\tablewidth{0pt}
\tablehead{
  \raisebox{-2ex}[0pt]{DiRB status}&
  \multicolumn{2}{c}{Synchrotron bump}&&
  \multicolumn{3}{c}{Inverse Compton bump}
  \\
  \cline{2-3}
  \cline{5-7}
  \colhead{}&
  \colhead{$\log_{10} \nu_s^{\max}$}&
  \colhead{$\log_{10} \left( \nu_s \frac{\D F_s^{}}{\D \nu_s^{}} \right)_{\max}$ }&
  \colhead{}&
  \colhead{$\log_{10} \nu_c^{\max}$ }&
  \colhead{$\log_{10} \left( \nu_c \frac{\D F_c^{}}{\D \nu_c^{}} \right)_{\max}$}&
  \colhead{$r_{\rm max}$ }
}
\startdata
%%%%%%%%%%%%%%%%%%%%%%%%%%%%%%%%%%%%%%%%%%%%%%%%%%%%%%%%%%%%%%%%%%%%%%
reddened&&&& $26.4\pm0.1$ & $-9.46\pm0.04$ & 0.4467 \\
de-reddened&\raisebox{1.5ex}[0pt]{$19.3\pm0.1$}  & 
\raisebox{1.5ex}[0pt]{$-9.11\pm0.04$} && 
  $26.6\pm0.1$ & 
 $-9.04\pm0.04$ & 
 1.175 
%%%%%%%%%%%%%%%%%%%%%%%%%%%%%%%%%%%%%%%%%%%%%%%%%%%%%%%%%%%%%%%%%%%%%%
\enddata
\end{deluxetable}
\clearpage
\begin{deluxetable}{ccccc}
\tablecaption{%%	
   Values of function $\alpha_{\gamma \gamma}(\beta)$ and the modified 
   function modified function $\tilde\alpha_{\gamma \gamma}(\beta)$ 
   as function as the spectral index $\beta$ in $\nu F_\nu$.
   \label{tab:alphatau}
}
\tablewidth{0pt}
\tablehead{ 
  \colhead{$\beta$} &
  \colhead{0} &
  \colhead{$\frac{1}{2}$} & 
  \colhead{1} &
  \colhead{$\frac{4}{3}$} 
}
%%%%%%%%%%%%%%%%%%%%%%%%%%%%%%%%%%%%%%%%%%%%%%%%%%%%%%%%%%%%%%%%%%%%%%
\startdata
  $\alpha_{\gamma \gamma}(\beta)$ 
    & 0.122 
    & 0.236 
    & 0.583 
    & 1.397\\
  $\tilde{\alpha}_{\gamma \gamma} ( \beta )$ 
    & -- 
    & 0.043 
    & 0.583 
    & 1.613
\enddata
\end{deluxetable}
\clearpage
\begin{figure}
  \plotone{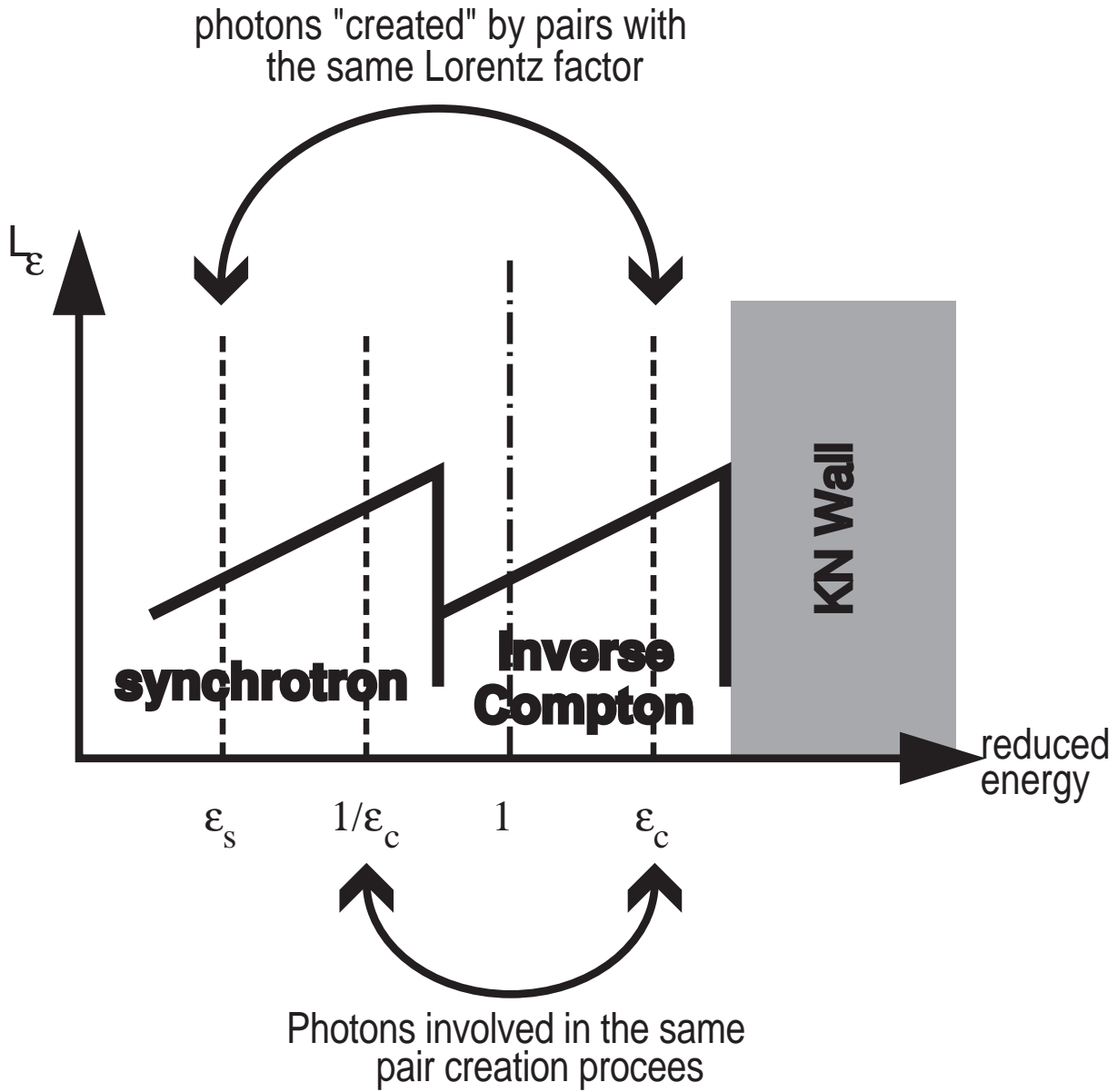}
  \caption{%%
    Working figure -- High energy photons with reduced energy $\varepsilon_c^{\ast}$ interact
    preferentialy with soft photons of energy $1 / \varepsilon_c^{\ast}$ to
    create new pairs. In the Klein-Nishina scatering regime, an
    ultrarelativistic particle of reduced energy $\gamma =
    \mbox{$\varepsilon_c^{\ast}$}$ "create" a high energy photon with the
    same energy $\varepsilon_c^{\ast}$, and in the same time, other pairs
    with the same energy can create soft photons by synchroton process of
    energy $\varepsilon_s^{\ast} = ({3 e h}/{4 \pi m_e^2 c^3})\, B \gamma^2
    = ({B}/{B_0})\, \gamma^2$.
  }\label{fig1}
\end{figure}
\clearpage
\begin{figure}
\plotone{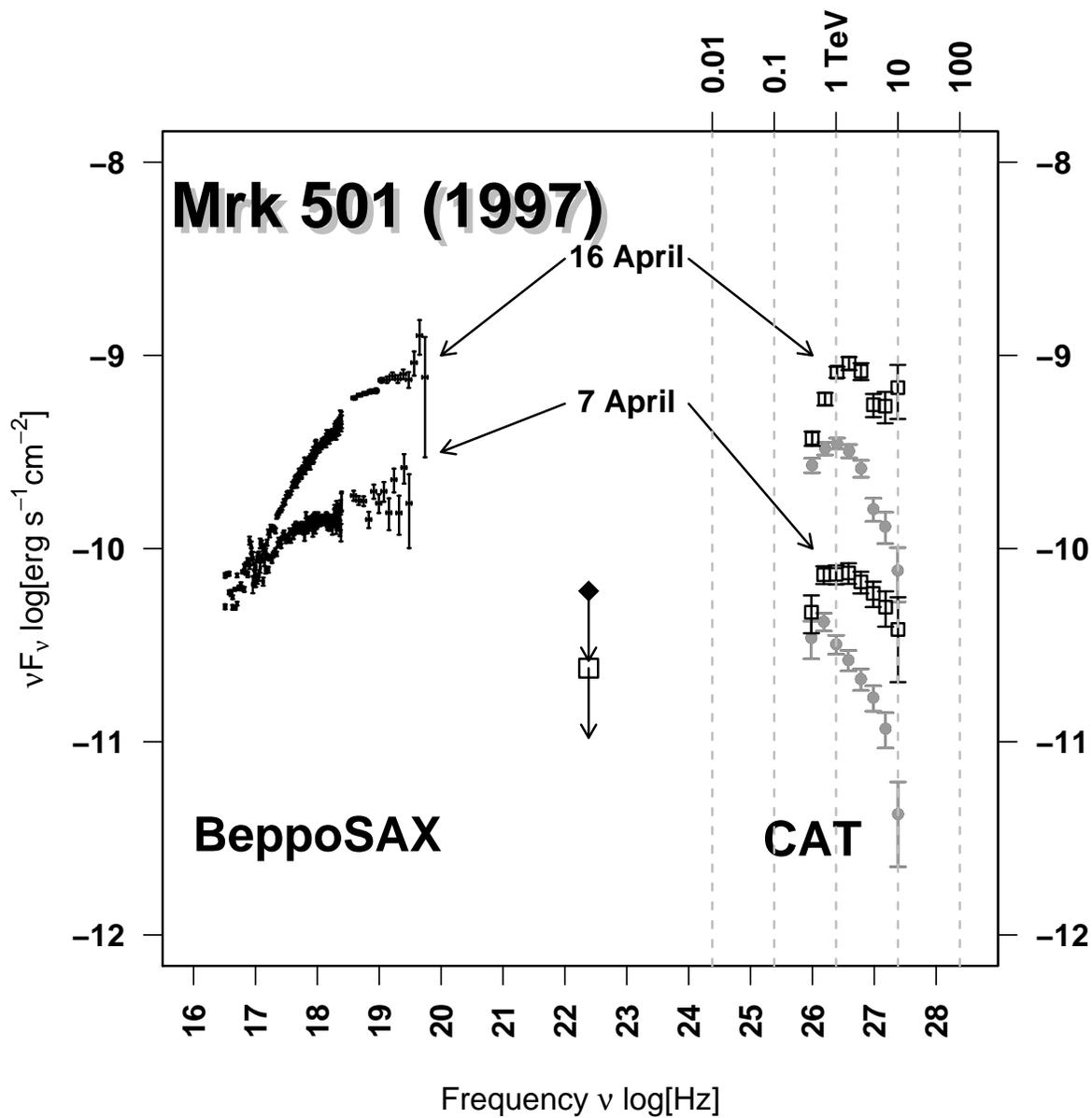}
\caption{% 
  Spectral energy distribution of Mrk 501 during the flaring period in
  1997 April, showing the simultaneous data taken by the Beppo-SAX
  instrument \citep{pian98} and by the CAT imaging Atmospheric
  {\vC}erenkov Telescope  \citep{cat99,barrau98}. About high energy data
  points, filled gray circles are the CAT observed ones while open
  squares are unabsorbed ones, corrected from our estimation of the
  DIrB attenuation%%
}\label{fig2}
\end{figure}
\clearpage
\begin{figure}
\plottwo{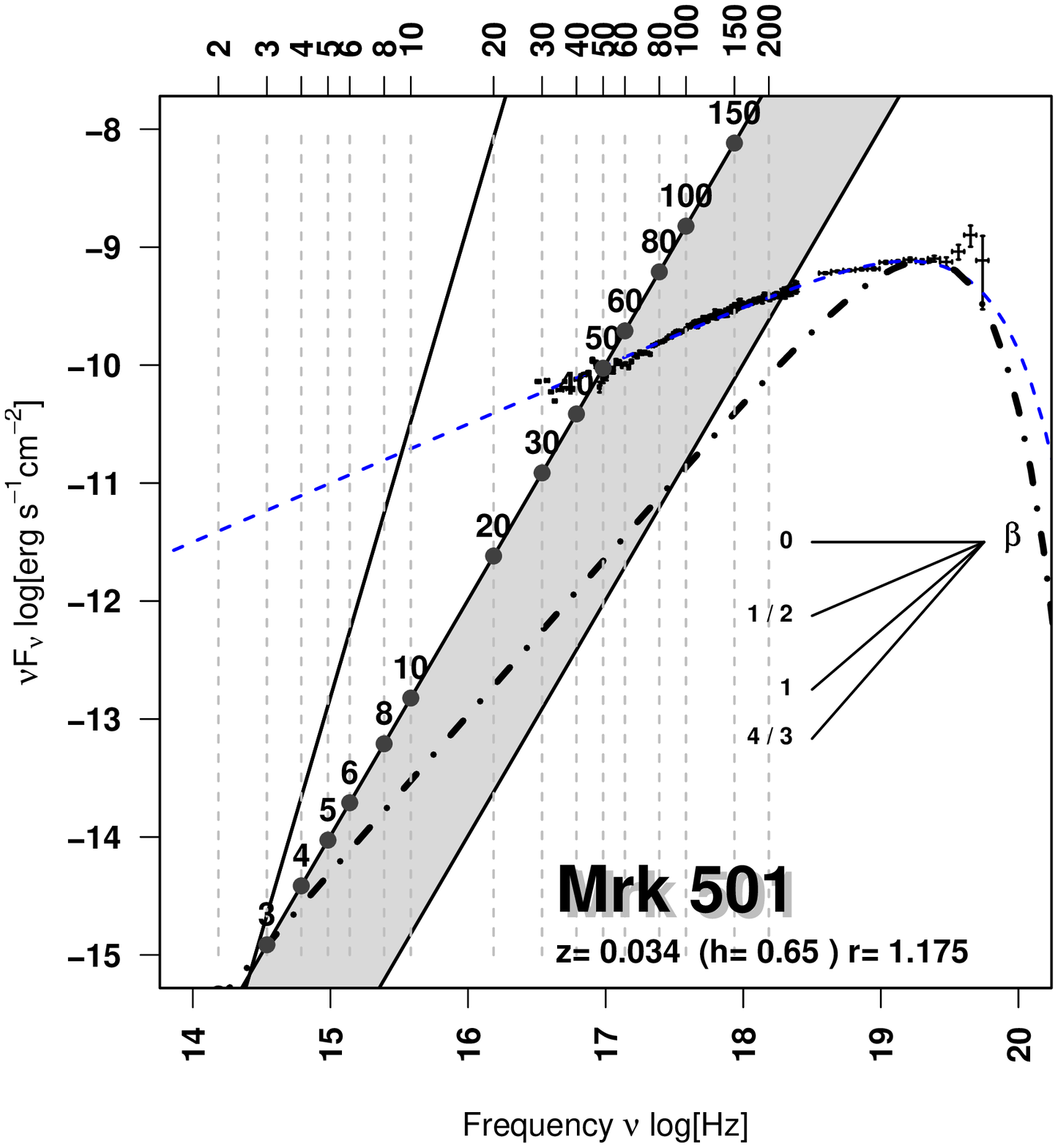}{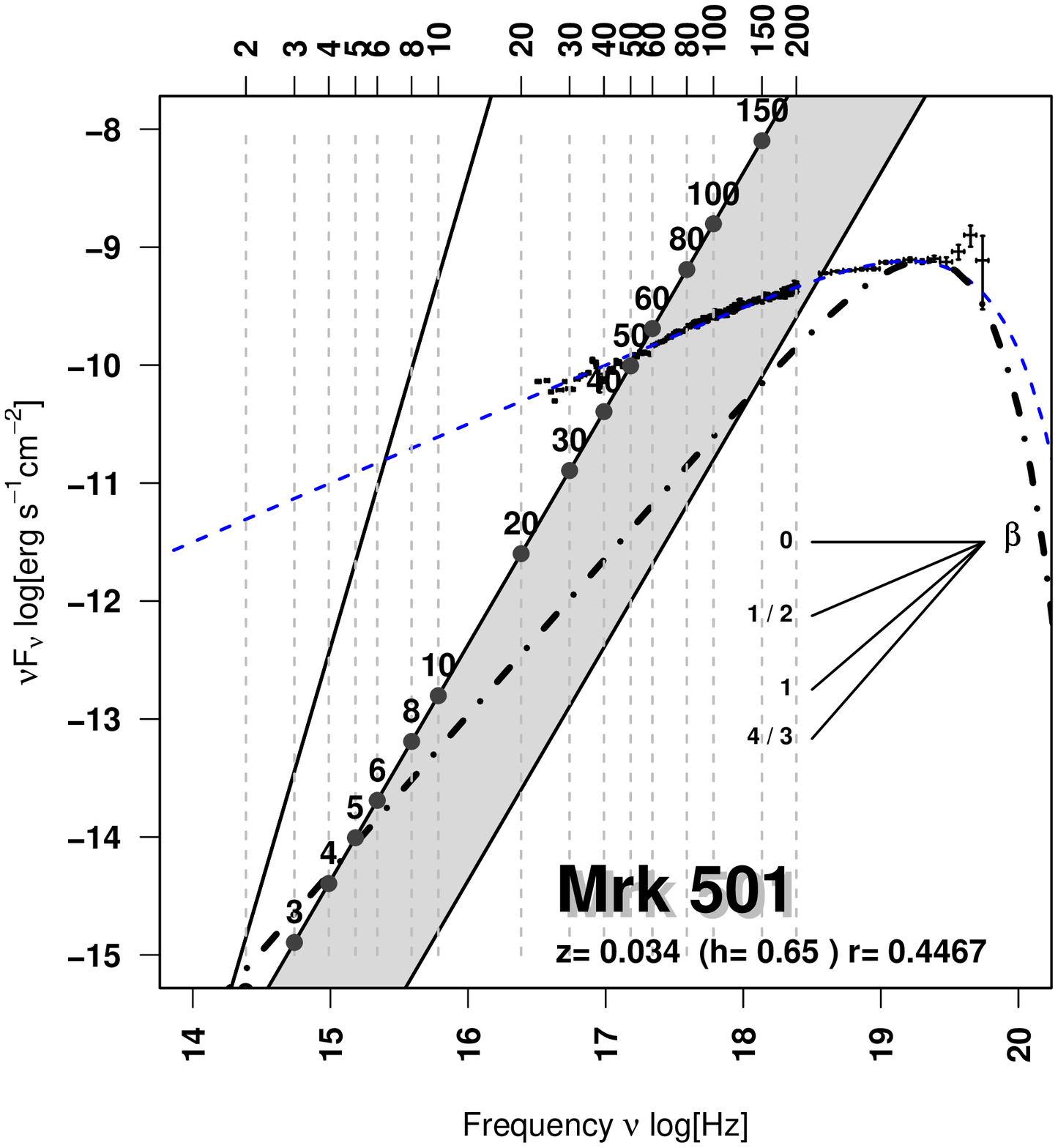}
\caption{%
        Constraints on the local shape of the synchrotron spectrum of
        Mrk~501 during the high 1997 April 16 high state. Grayed polygon is
        obtained considering the gamma transparency argument. It is defined
        by the zone where opacity lying into the interval
        $\tau_{\gamma\gamma}\times A^{1/2}\in[0.1,1]$ where $A$ is the
        aspect ratio. For an homogeneous spherical blob $A=1$ while in the
        case of a jet $A=1/\theta_j$ where $\theta_j$ is the characteristic
        opening angle of the jet. Constraint coming from the typical
        variability time scale leads to the most left straight thick line.
        Also represented in dotted-dashed line, a spectrum with a spectral
        index equals to $4/3$ in $\nu F_{\nu}$ resulting from the emission
        of a (quasi-)monoenergetic distribution of electrons and/or
        positrons. Left (resp. right) pannel correspond to the de-reddened
        (resp. reddened) case.  }\label{fig34}
\end{figure}
\clearpage
\begin{figure}
\plotone{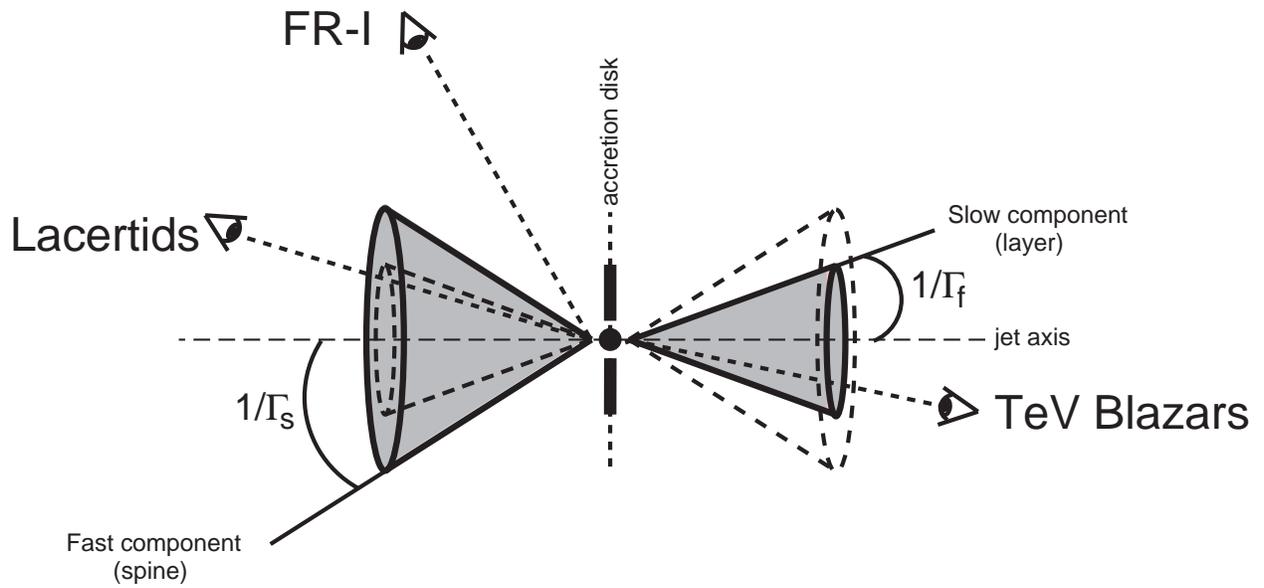}
\caption{%
    Sketch of the Two pattern model. It considers a relativistic external layer
    characterized by a Lorentz factor $\Gs$ and a fast internal spine with
    $\Gs>\Gf$. Considering relativistic Doppler beaming, an object viewed with
    $\theta>1/\Gs$ refers to a FR-I radio galaxy. Conversely, if
    $\theta\leqslant1/\Gs$ the source is seen as a BL Lac object and more
    precisely, if $\theta\leqslant1/\Gf$ the fast inner component dominate the
    emission. In this latter case one deals with a TeV blazar.  }\label{fig5}
\end{figure}
\clearpage
\begin{figure}
\plotone{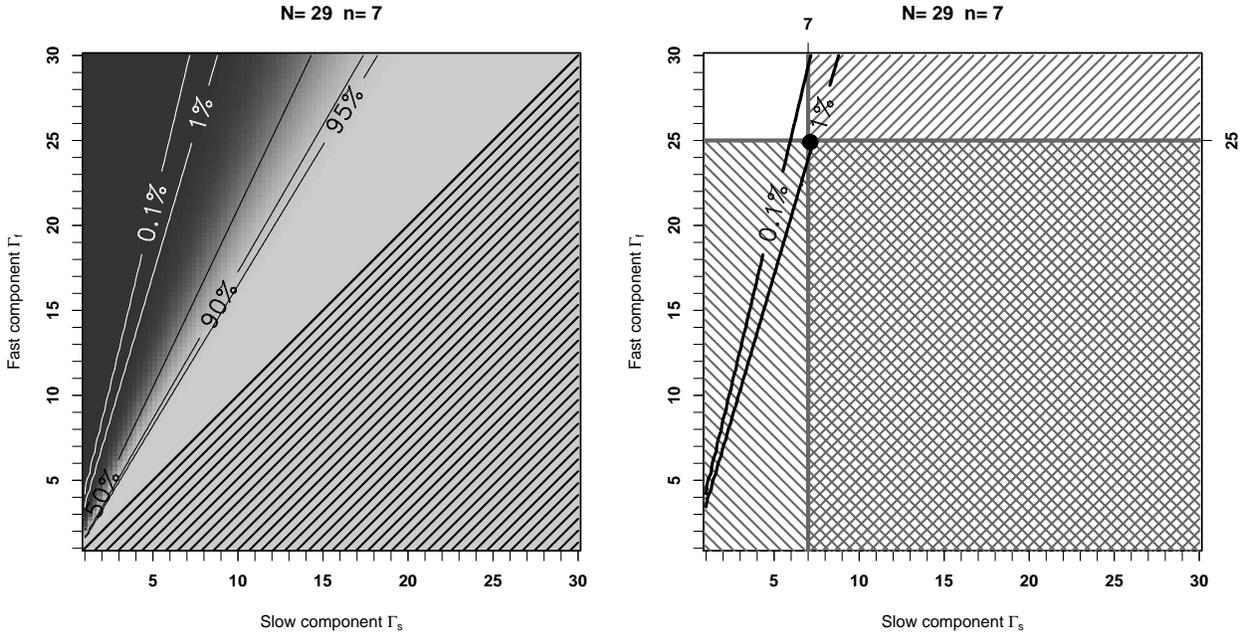}
\caption{%
    \emph{Left panel.} Space parameters $(\Gs,\Gf)$ constrained by the
    statistics of the number $n$ of TeV blazars in a given population of $N$
    observed Lacertids.  Each line  corresponds equivelently to a constant
    value of $\mathcal P_0=\mathcal P(29\geqslant n_0\geqslant 7)$ or of the
    ratio $\Gs/\Gf$. Here represented $\mathcal P_0={95\%, 90\%, 50\%,
    1\%\mbox{ and } 0.1\%}$ (see text for more details).\ \emph{Right panel.}
    Same as previous panel but combined with the constraints coming from
    $\gamma-$rays transparency argument and FR-I/Lacertids unification models.
    The first one eliminates all the region lying above the $\Gf{}_{,\min}=25$
    while the seconde one suppress the right part of the parameters space
    $\Gs\geqslant\Gs{}_{,\max}=7$. In this case, the allowed region compatible
    with a probability of detection of at least 7 TeV blazars in a population
    29 Lacertids is strongly improbable.  }\label{fig6}
\end{figure}
\end{document}